\documentclass[12pt]{article}
\usepackage{epsfig}

\def\beq{\begin{equation}}
\def\eeq#1{\label{#1}\end{equation}}
\def\eeqn{\end{equation}}

\def\beqa{\begin{eqnarray}}
\def\eeqa#1{\label{#1}\end{eqnarray}}
\def\eeqan{\end{eqnarray}}

\let\bar=\overbar

\def\Dslash{\not{\hbox{\kern-4pt $D$}}}
\def\dslash{\not{\hbox{\kern-2pt $\del$}}}

\def\msb{{\bar{\ssstyle M \kern -1pt S}}}

\def\Title#1{\begin{center} {\Large {\bf #1} } \end{center}}

\begin{document}

\Title{Higgs phenomenology beyond the Standard Model} 

\bigskip\bigskip


\begin{raggedright}  

{\it Heather E.\ Logan\index{Logan, H.\ E.}\\
Ottawa-Carleton Institute for Physics\\
Carleton University\\
Ottawa K1S 5B6 CANADA}
\bigskip\bigskip
\end{raggedright}

\centerline{\bf Abstract}
Detection of a signal in one of the standard LHC Higgs search channels
does not guarantee that the particle discovered is the Standard Model (SM)
Higgs.  In this talk I survey some general classes of alternatives and ways to
tell them apart.

\section{Introduction}

A non-SM Higgs boson may manifest in a variety of ways.  In this talk I focus on the production and decay of a non-SM Higgs in SM Higgs search channels.
In this case the ``non-SM-ness'' of the Higgs manifests experimentally via modified couplings to SM particles.  There are three ways in which such modified couplings typically arise.  First, new particles running in the loop can modify the loop-induced cross section for $gg \to H$ or the partial width for $H \to \gamma\gamma$.  Second, when two or more mass eigenstates share the electroweak symmetry breaking (EWSB) vacuum expectation value (vev), the couplings of either eigenstate to $WW$, $ZZ$ are modified.  This modifies the Higgs production cross sections in vector boson fusion and in $WH$, $ZH$ associated production, as well as modifying the partial widths for $H \to WW$ and $H \to ZZ$.  It also modifies the partial width for $H \to \gamma\gamma$ due to the contribution of the $W$ boson in the loop.  Third, in models with more than one Higgs doublet, the masses of different fermions can arise from different Higgs doublets.  This can modify the cross section for $gg \to H$ by shifting the Higgs coupling to $t \bar t$.  It can also change the ratios of Higgs partial widths to different fermion species, and can affect all Higgs branching fractions (including to non-fermionic final states) by modifying the Higgs total width.
In what follows I survey the effects of these coupling modifications in a selection of non-SM Higgs models and point out prospects for their detection at the LHC.  I finish with a review of techniques to extract individual Higgs couplings by combining LHC measurements in different channels, including a new idea using the Higgs total width.

\section{Higgs signals in non-SM models}

{\it Fourth-generation model:}
If a fourth sequential generation of fermions exists, the new heavy fermions running in loops substantially affect $gg \to H$ and $H \to \gamma\gamma$.  The $gg \to H$ amplitude is independent of the quark mass in the limit $m_q \gg M_H$; the two heavy fourth-generation quarks add to the top quark to essentially triple the amplitude, resulting in an enhancement of the $gg \to H$ cross section by a factor of nine.  The $H \to \gamma\gamma$ partial width, on the other hand, is generically suppressed due to destructive interference between the new $t^{\prime}$, $b^{\prime}$, and $\tau^{\prime}$ loops and the SM $W$ loop.  Higgs rates at the LHC in various channels were studied in Ref.~\cite{Gunion:2011ww}.  After this conference, ATLAS and CMS presented updated Higgs search results that exclude fourth-generation model Higgs masses between 120 and 600~GeV~\cite{ATLASHiggsLP,CMSHiggsEPS}.

{\it Minimal Supersymmetric SM:}
Couplings of the light MSSM Higgs $h^0$ are modified by squarks and charginos running in the $gg \to h^0$ and $h^0 \to \gamma\gamma$ loops, by the sharing of the EWSB vev between two mass eigenstates $h^0$ and $H^0$, and by the nontrivial coupling pattern of the two Higgs doublets to fermions.  These effects are most pronounced when all the Higgs states are relatively light, and can lead to a suppression of the important $gg \to h^0 \to \gamma\gamma$ channel.  This can lead to a ``hole'' in the LHC MSSM Higgs coverage at low pseudoscalar mass $M_A \leq 150$~GeV and moderate to large $\tan\beta \geq 10$~\cite{Carena:2011fc}, which could be compensated by using the complementary $W/Zh^0$, $h^0 \to b \bar b$ channel at the Tevatron.  This difficult region has subsequently been mostly excluded by direct searches for $A^0,H^0 \to \tau\tau$~\cite{CMSmssmPIC,ATLASmssmLP}, illustrating the huge importance of the interplay among different search channels in specific models.

{\it More general two-Higgs-doublet models:}
Two-Higgs-doublet models (2HDMs) have similar features to the MSSM Higgs sector with in general fewer theoretical constraints.  They also open the possibility for different fermion coupling structures.  For example, in the ``lepton-specific'' 2HDM, one Higgs doublet couples to quarks while the other couples to leptons, allowing for suppression of Higgs couplings to quarks while couplings to leptons are enhanced.  The effects on Higgs rates in the LHC search channels have been studied, e.g., in Ref.~\cite{Su:2009fz}.  Similarly, in the ``flipped'' 2HDM, one doublet couples to up-type quarks and charged leptons while the other couples to down-type quarks, allowing for a large Higgs coupling to $b\bar b$ while suppressing Higgs decays to $\tau\tau$.

{\it Top-Higgs:}
The top-Higgs is an add-on for models of dynamical EWSB.  It introduces a dedicated Higgs doublet (generally composite) to generate most of the top quark mass.  The top-Higgs particle $H_T$ couples only to $t\bar t$, $WW$, and $ZZ$ at tree level with a distinctive coupling pattern: writing the top-Higgs doublet vev $f = v_{\rm SM} \sin\omega$, the couplings to $WW/ZZ$ are suppressed by $\sin\omega$ while the coupling to $t \bar t$ is enhanced by $1/\sin\omega$, leading to an enhancement of the $gg \to H_T$ cross section by $1/\sin^2\omega$.  Tevatron limits on $gg \to H \to WW$~\cite{Aaltonen:2010sv} already exclude a range of parameter space around $M_{H_T} \sim 150$--220~GeV~\cite{Chivukula:2011ag}; more recent ATLAS and CMS Higgs searches~\cite{ATLASHiggsLP,CMSHiggsLP} exclude most of the favoured top-Higgs parameter space between 200 and 350~GeV~\cite{Chivukula:2011dg}.

{\it Lee-Wick Standard Model:}
The Lee-Wick SM~\cite{Grinstein:2007mp} is an exotic approach to solve the hierarchy problem by implementing Pauli-Villars regularization with actual physical fields.  Each SM field has a partner with wrong-sign quadratic Lagrangian terms.  The most interesting feature in the Higgs sector is the novel hyperbolic mixing structure between the SM Higgs $h$ and its Lee-Wick partner $\tilde h$:
\begin{equation}
	\left( \begin{array}{c} h \\ \tilde h \end{array} \right) = 
	\left( \begin{array}{cc} \cosh\theta & \sinh\theta \\
		\sinh\theta & \cosh\theta \end{array} \right)
	\left( \begin{array}{c} h_0 \\ \tilde h_0 \end{array} \right).
\end{equation}
The usual sum rule for CP-even neutral Higgs couplings to $WW$ (normalized to the SM), $\bar g_{hWW}^2 + \bar g_{HWW}^2 = 1$, becomes $\bar g_{h_0WW}^2 - \bar g_{\tilde h_0WW}^2 = 1$, leading to enhancement of the $h_0WW$ coupling above its SM value, a feature inaccessible in ordinary 2HDMs.  Higgs couplings to fermions are also enhanced.  Tevatron exclusions and LHC prospects were studied in Ref.~\cite{Alvarez:2011ah}.  Based on those results, the most recent LHC SM Higgs exclusions~\cite{ATLASHiggsLP,CMSHiggsLP} should exclude masses above $\sim$145~GeV for the SM-like Higgs $h_0$.

\section{Higgs coupling extraction}

To test the Higgs mechanism and characterize extended Higgs models, we need to measure the Higgs couplings.  Individual Higgs couplings can be extracted at the LHC by combining measurements of rates in different channels~\cite{Zeppenfeld:2000td,Belyaev:2002ua,Duhrssen:2004cv,Lafaye:2009vr}.  This fit possesses a flat direction corresponding to allowing an unobserved decay mode with width $\Gamma_{\rm new}$ while simultaneously increasing all couplings by a factor of $a$ so as to leave all observed signal rates unchanged.
While no well-motivated models exhibit this behaviour, it precludes a truly model-independent extraction of the Higgs couplings.  This has usually been dealt with by either assuming no non-SM decays are present~\cite{Zeppenfeld:2000td,Lafaye:2009vr} or assuming that the Higgs couplings to $WW$, $ZZ$ are bounded from above by their SM values~\cite{Duhrssen:2004cv}; the latter is true in most models (the Lee-Wick SM and models with SU(2) triplets being exceptions).  An alternate approach, useful for SM Higgs masses above $\sim$190~GeV, is to incorporate the Higgs width measurement from the $H \to ZZ \to 4\ell$ lineshape in the coupling fit~\cite{Logan:2011ey}.  For a SM Higgs with mass 190~GeV, this approach yields uncertainties of about 10\% on the Higgs couplings-squared to $WW$, $ZZ$, and $gg$, and a 2$\sigma$ limit on $\Gamma_{\rm new}$ of about $0.25 \Gamma_{\rm tot}^{\rm SM}$ (using 30~fb$^{-1}$ at one detector at 14~TeV).  While a 190~GeV SM Higgs has now been excluded by LHC data~\cite{ATLASHiggsLP,CMSHiggsLP}, this method remains applicable to scenarios in which the Higgs is heavier and/or signal rates in SM channels are suppressed.

\section{Summary}

With the excellent operation of the LHC throughout 2011, discovery or exclusion of a SM Higgs is imminent.  A signal in one of the SM Higgs channels will have immediate impact on non-SM Higgs scenarios as well.  Higgs coupling measurements will then be key to understanding the structure of the Higgs sector.

\bigskip
I thank the organizers of Physics at the LHC 2011 for the invitation to speak, and for providing such a pleasant and inspiring environment.  This work was supported by the Natural Sciences and Engineering Research Council of Canada.

\end{document}